\documentclass[12pt]{article}

\usepackage{amsfonts}
\usepackage{amsmath,amssymb,amsthm}
\usepackage{appendix}
\usepackage{bbm} 
\usepackage{amsbsy}
\usepackage{enumerate}
\usepackage{cite}
\usepackage{enumerate}
\usepackage[dvips]{graphicx}
\usepackage{url}
\usepackage{hyperref}
\hypersetup{colorlinks,citecolor=blue,linkcolor=blue,urlcolor=black}

\usepackage{bbm}
\newcommand{\indic}{\mathbbm{1}}
\newcommand{\ind}[1]{\indic_{\{#1\}}}

\numberwithin{equation}{section}
\theoremstyle{plain}

\newtheorem{thm}{Theorem}[section]

\newtheorem{lmm}[thm]{Lemma}

\newcommand{\cD}{{\cal D}}

\newcommand{\conn}{\longleftrightarrow}

\newcommand{\dc}{d_{\rm c}}
\newcommand{\ds}{d_{\rm short}}

\newcommand{\dpst}{\displaystyle}
\newcommand{\etas}{\eta_{\rm short}}

\newcommand{\lbeq}[1]{\label{eq:#1}}

\newcommand{\mP}{{\mathbb P}}
\newcommand{\mR}{{\mathbb R}}
\newcommand{\mZ}{{\mathbb Z}}
\newcommand{\N}{\mathbb{N}}
\newcommand{\nn}{\nonumber}
\newcommand{\pc}{p_{\rm c}}

\newcommand{\Rd}{\mR^d}
\newcommand{\refeq}[1]{(\ref{eq:#1})}
\newcommand{\sss}{\scriptscriptstyle}
\newcommand{\veee}[1]{|\hskip-1.4pt|\hskip-1.4pt|#1|\hskip-1.4pt|\hskip-1.4pt|}
\newcommand{\Zd}{\mZ^d}

\title{Crossover phenomena in the critical behavior for long-range 
models with power-law couplings}

\author{
Akira~Sakai\footnote{Faculty of Science, Hokkaido University, Japan. 
\url{https://orcid.org/0000-0003-0943-7842}}
}

\begin{document}
\maketitle

\begin{abstract}
This is a short review of the two papers \cite{csIV,csV} on the $x$-space 
asymptotics of the critical two-point function $G_{\pc}(x)$ for the long-range 
models of self-avoiding walk, percolation and the Ising model on $\Zd$, defined 
by the translation-invariant power-law step-distribution/coupling 
$D(x)\propto|x|^{-d-\alpha}$ for some $\alpha>0$.  Let $S_1(x)$ be the 
random-walk Green function generated by $D$.  We have shown that
\begin{itemize}
\item
$S_1(x)$ changes its asymptotic behavior from Newton ($\alpha>2$) to Riesz 
($\alpha<2$), with log correction at $\alpha=2$;
\item
$G_{\pc}(x)\sim\frac{A}{\pc}S_1(x)$ as $|x|\to\infty$ in dimensions higher than 
(or equal to, if $\alpha=2$) the upper critical dimension $\dc$ (with 
sufficiently large spread-out parameter $L$).  The model-dependent $A$ and 
$\dc$ exhibit crossover at $\alpha=2$.  
\end{itemize}
The keys to the proof are (i)~detailed analysis on the underlying random walk 
to derive sharp asymptotics of $S_1$, (ii)~bounds on convolutions of power 
functions (with log corrections, if $\alpha=2$) to optimally control the 
lace-expansion coefficients $\pi_p^{\sss(n)}$, and (iii)~probabilistic 
interpretation (valid only when $\alpha\le2$) of the convolution of $D$ and 
a function $\varPi_p$ of the alternating series 
$\sum_{n=0}^\infty(-1)^n\pi_p^{\sss(n)}$.  We outline the proof, emphasizing 
the above key elements for percolation in particular.
\end{abstract}

\section{Introduction and the main results}
Since the dawn of research on phase transitions and critical behavior, 
it has been standard to investigate short-range models, among which 
the nearest-neighbor model on $\Zd$ is the most popular.  Thanks to 
intensive studies for more than half a century, nearest-neighbor bond 
percolation is now known to exhibit a phase transition for all $d\ge2$ 
and mean-field behavior (i.e., the critical two-point function $G_{\pc}(x)$ 
decays as $|x|^{2-\etas-d}$ with the mean-field value $\etas=0$) for all 
$d\ge11$ \cite{fh17a,fh17b}.  Believing in universality, we expect the 
mean-field behavior for all dimensions above the upper-critical dimension 
$\ds=6$ for short-range percolation \cite{hs90p}.

Recently, long-range random walk and statistical-mechanical models defined 
by the power-law step-distribution/coupling $D(x)\propto|x|^{-d-\alpha}$, 
$\alpha>0$, have regained popularity, due to unconventional macroscopic 
behavior \cite{bpr14,csI,csII,csIII,csIV,csV,hhs08,lsw17}.  Among those 
references, an infrared bound and mean-field behavior are proven for 
long-range oriented percolation (OP for short) on $\mZ^{d>\dc}\times\mZ_+$ 
\cite{csI,csII} and for long-range models of percolation, self-avoiding walk 
(SAW for short) and the Ising model on $\mZ^{d>\dc}$ \cite{hhs08}, where
\begin{align}\lbeq{dcdef}
\dc=(\alpha\wedge2)\times
 \begin{cases}
 2\quad&[\text{OP, SAW \& Ising}],\\
 3&[\text{percolation}].
 \end{cases}
\end{align}
Also, an asymptotic expression of the gyration radius for long-range models of 
SAW and OP for $d>\dc$ are proven in \cite{csIII}.  In physics, Brezin, Parisi 
and Ricci-Tersenghi \cite{bpr14} conjectured that $G_{\pc}(x)$ would decay as 
$|x|^{\alpha\wedge(2-\etas)-d}$ if $\alpha\ne2-\etas$, and as 
$|x|^{\alpha-d}/\log|x|$ if $\alpha=2-\etas$.  We have shown in \cite{csIV,csV} 
that the conjectured behavior holds true for $d>\ds$ ($=\dc$ with 
$\alpha=\infty$), because $\etas=0$, with sufficiently large spread-out 
parameter $L$ \cite{h08,hhs03,s07}.  In fact, the obtained results are much 
stronger, as summarized as follows.

\begin{thm}[Proposition~2.1 of \cite{csIV} and Theorem~1.3 of \cite{csV}]
\label{theorem:S}
Let $\alpha>0$, $L\ge1$ and 
$D(x)\asymp\tfrac1{L^d}(\frac{|x|}L\vee 1)^{-d-\alpha}$, i.e., 
\begin{align}\lbeq{Ddef}
{}^\exists c>0,~{}^\forall x\in\Zd,~{}^\forall L\in[1,\infty):~c\le\frac{D(x)}
 {\frac1{L^d}(\frac{|x|}L\vee 1)^{-d-\alpha}}\le\frac1c.
\end{align}
Let
\begin{align}\lbeq{gamma}
\gamma_\alpha=\frac{\Gamma(\frac{d-\alpha\wedge2}2)}{2^{\alpha\wedge2}\pi^{d/2}
 \Gamma(\frac{\alpha\wedge2}2)},&&&&
v_\alpha=
 \begin{cases}
 \dpst\lim_{|k|\to0}\frac{1-\hat D(k)}{|k|^{\alpha\wedge2}}&[\alpha\ne2],\\
 \dpst\lim_{|k|\to0}\frac{1-\hat D(k)}{|k|^2\log(1/|k|)}\quad&[\alpha=2],
 \end{cases}
\end{align}
where $\hat D(k)=\sum_{x\in\Zd}e^{ik\cdot x}D(x)$.
Then, for all $d>\alpha\wedge2$, the random-walk Green function $S_1(x)$ 
generated by the step distribution $D$ exhibits the following asymptotic 
behavior: there is an $\epsilon>0$ such that, as $|x|\to\infty$,
\begin{align}\lbeq{S1asy}
S_1(x)=\frac{\gamma_\alpha/v_\alpha}{|x|^{d-\alpha\wedge2}}\times
 \begin{cases}
 \dpst\bigg(1+\frac{O(L^{\epsilon})}{|x|^\epsilon}\bigg)&[\alpha\ne2],\\[1pc]
 \dpst\frac1{\log|x|}\bigg(1+\frac{O(1)}{(\log|x|)^\epsilon}\bigg)\quad
  &[\alpha=2],
 \end{cases}
\end{align}
where the $O(1)$ term is independent of $L$.
\end{thm}

\begin{thm}[Theorem~1.2 of \cite{csIV} and Theorem~1.6 of \cite{csV}]
\label{thm:main}
Let $D$ be the same as in Theorem~\ref{theorem:S} and recall the definition 
\refeq{dcdef} of $\dc$.  Suppose $d>\dc$ for $\alpha\ne2$ or that $d\ge\dc$ 
for $\alpha=2$.  For $\alpha>2$, we also assume a bound on the ``derivative" 
of $D$ (see the last part of Section~\ref{s:main}).  Then, there is an 
$L_0(d)<\infty$ such that, for any $L\ge L_0$, there are 
$A=1+O(L^{-2})\ind{\alpha>2}$ and $\epsilon>0$ such that, as $|x|\to\infty$,
\begin{align}\lbeq{mainasy}
G_{\pc}(x)=\frac{A}{\pc}\frac{\gamma_\alpha/v_\alpha}{|x|^{d-\alpha\wedge2}}
 \times
 \begin{cases}
 \dpst\bigg(1+\frac{O(L^\epsilon)}{|x|^\epsilon}\bigg)&[\alpha\ne2],\\[1pc]
 \dpst\frac1{\log|x|}\bigg(1+\frac{O(1)}
 {(\log|x|)^\epsilon}\bigg)\quad&[\alpha=2],
 \end{cases}
\end{align}
where the $O(1)$ term is independent of $L$.
\end{thm}

In short, the critical two-point function $G_{\pc}(x)$ exhibits the same 
asymptotic behavior as $S_1(x)$, modulo multiplication of the model-dependent 
constant $A/\pc$, for all $d>\dc$ (with large spread-out parameter $L$) and, 
most interestingly, for $d=\dc$ when $\alpha=2$.  For $d\in(\dc,\ds)$, which is 
not empty for $\alpha<2$ and in which $\etas$ is believed to be nonzero, 
Theorem~\ref{thm:main} claims that $G_{\pc}(x)$ decays as $|x|^{\alpha-d}$, 
not as $|x|^{2-\etas-d}$.  This power-law behavior has been 
extended even below $\dc$ by Lohmann, Slade and Wallace \cite{lsw17} using 
a rigorous version of the $\varepsilon$-expansion.

\section{Key ideas for the proof of Theorem~\ref{theorem:S}}
Let $D^{*n}$ be the $n$-fold convolution of $D$ (i.e., the $n$-step 
distribution) and denote by $S_q$ the random-walk Green function generated 
by $D$ with survival rate $q\in[0,1]$:
\begin{gather}
D^{*n}(x)=(D^{*(n-1)}*D)(x)\equiv\sum_yD^{*(n-1)}(y)\,D(x-y),\\
S_q(x)=\sum_{n=0}^\infty q^nD^{*n}(x).
\end{gather}
Let
\begin{align}\lbeq{veee}
\veee{x}_r=\frac\pi2(|x|\vee r)\qquad[x\in\Rd,~1\le r<\infty],
\end{align}
where $|\cdot|$ is the Euclidean norm.  Suppose that, as explained in 
\refeq{Ddef}, $D(x)$ decays as 
\begin{align}\lbeq{Dredef}
D(x)\asymp L^{-d}\veee{\tfrac{x}L}_1^{-d-\alpha}\equiv L^\alpha
 \veee{x}_L^{-d-\alpha}.
\end{align}
An example of $D$ is the following compound zeta distribution \cite{csIV}:
\begin{align}\lbeq{comp-zeta}
D(x)=\sum_{t\in\N}U_L^{*t}(x)\,\frac{t^{-1-\alpha/2}}{\zeta(1+\alpha/2)}
 \qquad[x\in\Zd],
\end{align}
where $U_L$ is the uniform distribution over the $d$-dimensional box 
of side-length $2L$.

The step distribution $D$ in \refeq{Dredef} satisfies the following properties 
(D1)--(D3) that are essential to the proof of \refeq{S1asy}.
\begin{enumerate}
\item[(D1)]
$k$-space bounds \cite[Proposition~1.1]{csI} (and \cite[Assumption~1.1]{csV}): 
${}^\exists\Delta=\Delta(L)\in(0,1)$ such that
\begin{align}\lbeq{1-hatDbd1}
1-\hat D(k)
 \begin{cases}
 <2-\Delta\quad&[{}^\forall k\in[-\pi,\pi]^d],\\
 >\Delta&[|k|>1/L],
 \end{cases}
\end{align}
and for $|k|\le1/L$,
\begin{align}\lbeq{1-hatDbd2}
1-\hat D(k)\asymp (L|k|)^{\alpha\wedge2}\times
 \begin{cases}
 1&[\alpha\ne2],\\
 \log\frac\pi{2L|k|}\quad&[\alpha=2].
 \end{cases}
\end{align}
\item[(D2)]
$k$-space asymptotics \cite[Lemma~A.1]{csIII} (and \cite[Assumption~1.1]{csV}): 
${}^\exists\epsilon>0$ such that, as $|k|\to0$,
\begin{align}\lbeq{1-hatDasy}
1-\hat D(k)=v_\alpha|k|^{\alpha\wedge2}\times
 \begin{cases}
 \big(1+O(L^\epsilon|k|^\epsilon)\big)&[\alpha\ne2],\\
 \big(\log\frac1{L|k|}+O(1)\big)\quad&[\alpha=2],
 \end{cases}
\end{align}
where the constant in the $O(1)$ term is independent of $L$.
\item[(D3)]
$x$-space bounds \cite[(1.19)--(1.21)]{csIV} (and \cite[Assumption~1.2]{csV}): 
${}^\forall n\in\N$ and ${}^\forall x\in\Zd$,
\begin{align}
\|D^{*n}\|_\infty&\le O(L^{-d})\times
 \begin{cases}
 n^{-d/(\alpha\wedge2)}&[\alpha\ne2],\\
 (n\log\frac{\pi n}2)^{-d/2}\quad&[\alpha=2],
 \end{cases}\lbeq{Dnsupbd}\\
D^{*n}(x)&\le n\frac{O(L^{\alpha\wedge2})}{\veee{x}_L^{d+\alpha\wedge2}}\times
 \begin{cases}
 1&[\alpha\ne2],\\
 \log\veee{\frac{x}L}_1\quad&[\alpha=2].
 \end{cases}\lbeq{Dnxbd}
\end{align}
\end{enumerate}

For example, to show \refeq{1-hatDbd2} for $|k|\le1/L$, we first split the sum 
as 
\begin{align}
1-\hat D(k)\asymp L^\alpha\sum_x\veee{x}_L^{-d-\alpha}(1-\cos k\cdot
 x)~\Big(\ind{|x|<L}+\ind{L\le|x|\le\frac\pi{2|k|}}+\ind{|x|>\frac\pi{2|k|}}\Big).
\end{align}
It is easy to see that the contributions from the first and third indicators 
are $O(L^2|k|^2)$ and $O(L^\alpha|k|^\alpha)$, respectively.  The contribution 
from the second indicator is the main term since
\begin{align}
L^\alpha\sum_{L\le|x|\le\frac\pi{2|k|}}\veee{x}_L^{-d-\alpha}(1-\cos k\cdot x)
&\asymp L^\alpha|k|^2\sum_{L\le|x|\le\frac\pi{2|k|}}|x|^{-d-\alpha+2}\nn\\
&\asymp
 \begin{cases}
 (L|k|)^{\alpha\wedge2}&[\alpha\ne2],\\
 (L|k|)^2\log\frac\pi{2L|k|}&[\alpha=2].
 \end{cases}
\end{align}

To prove \refeq{S1asy}, we first rewrite $S_1(x)$ for the transient case 
$d>\alpha\wedge2$ as
\begin{align}
S_1(x)=\int_{[-\pi,\pi]^d}\frac{\text{d}^dk}{(2\pi)^d}\,\frac{e^{-ik\cdot x}}
 {1-\hat D(k)}&=\int_0^\infty\text{d}t\int_{[-\pi,\pi]^d}\frac{\text{d}^dk}
 {(2\pi)^d}\,e^{-ik\cdot x-t(1-\hat D(k))}\nn\\
&=\int_0^\infty\text{d}t\int_{|k|\le R}\frac{\text{d}^dk}{(2\pi)^d}\,e^{-ik
 \cdot x-t(1-\hat D(k))}+E_1,
\end{align}
where $R$ is arbitrary for the moment.  Then, by replacing $1-\hat D(k)$ by 
its limit \refeq{1-hatDasy}, we can further rewrite $S_1(x)$ for $\alpha\ne2$ as
\begin{align}
S_1(x)=\int_0^\infty\text{d}t\int_{\Rd}\frac{\text{d}^dk}{(2\pi)^d}\,e^{-ik\cdot
 x-v_\alpha t|k|^{\alpha\wedge2}}+E_1+E_2,
\end{align}
and for $\alpha=2$ as
\begin{align}
S_1(x)=\int_0^\infty\text{d}t\int_{\Rd}\frac{\text{d}^dk}{(2\pi)^d}\,e^{-ik\cdot
 x-v_2t|k|^2\log\frac1{L|k|}}+E_1+E_2.
\end{align}
Since
\begin{align}
\int_0^\infty\text{d}t\,e^{-v_\alpha t|k|^{\alpha\wedge2}}=\frac1{v_\alpha
 |k|^{\alpha\wedge2}}=\frac1{v_\alpha\Gamma(\frac{\alpha\wedge2}2)}
 \int_0^\infty\frac{\text{d}t}t\,t^{(\alpha\wedge2)/2}e^{-t|k|^2},
\end{align}
we readily obtain for $\alpha\ne2$ that
\begin{align}
S_1(x)-E_1-E_2=\frac1{v_\alpha\Gamma(\frac{\alpha\wedge2}2)}\int_0^\infty
 \frac{\text{d}t}t\,t^{(\alpha\wedge2)/2}\underbrace{\int_{\Rd}\frac{\text{d}^d
 k}{(2\pi)^d}\,e^{-ik\cdot x-t|k|^2}}_{=\,(4\pi t)^{-d/2}\exp(-|x|^2/(4t))}
 =\frac{\gamma_\alpha/v_\alpha}{|x|^{d-\alpha\wedge2}}.
\end{align}
Using the $k$-space and $x$-space bounds (D1) and (D3) and choosing $R$ 
accordingly (as in \cite[(2.20)]{csIV}), we can show that $E_1+E_2$ is the 
error term in \refeq{S1asy}.  See \cite[Section~2.1]{csIV} for more details.  

For $\alpha=2$, we change variables as $\xi=x/|x|$, $\kappa=|x|k$ and 
$\tau=\frac{v_2t}{|x|^2}\log\frac{|x|}L$ to obtain
\begin{align}
S_1(x)-E_1-E_2&=|x|^{-d}\int_0^\infty\text{d}t\int_{\Rd}\frac{\text{d}^d\kappa}
 {(2\pi)^d}\,\exp\bigg(-i\kappa\cdot\xi-\frac{v_2t|\kappa|^2}{|x|^2}
 \log\frac{|x|}{L|\kappa|}\bigg)\nn\\
&=\frac{|x|^{2-d}}{v_2\log\frac{|x|}L}\int_0^\infty\text{d}\tau\int_{\Rd}
 \frac{\text{d}^d\kappa}
 {(2\pi)^d}\,\exp\Bigg(-i\kappa\cdot\xi-\tau|\kappa|^2\frac{\log
 \frac{|x|}{L|\kappa|}}{\log\frac{|x|}L}\Bigg)\nn\\
&=\frac{|x|^{2-d}}{v_2\log\frac{|x|}L}\underbrace{\int_0^\infty\text{d}\tau
 \int_{\Rd}\frac{\text{d}^d\kappa}{(2\pi)^d}\,e^{-i\kappa\cdot\xi-\tau
 |\kappa|^2}}_{=\,\gamma_2}+E_3.
\end{align}
Again, by using the $k$-space and $x$-space bounds on $D$ and choosing $R$ 
accordingly (as in \cite[(2.5)]{csV}), we can show that $E_1+E_2+E_3$ is the 
error term in \refeq{S1asy}.  See \cite[Section~2.1]{csV} for more details.  
This completes the sketch proof of Theorem~\ref{theorem:S}.

\section{Key ideas for the proof of Theorem~\ref{thm:main}}\label{s:main}
The proof of Theorem~\ref{thm:main} is based on the lace expansion, which is 
one of the few methods to prove mean-field results mathematically rigorously.  
Since its invention by Brydges and Spencer for weakly SAW \cite{bs85}, 
the method has been extended to strictly SAW \cite{hs92}, oriented/unoriented 
percolation \cite{hs90p,ny93}, lattice trees and lattice animals \cite{hs90l}, 
the contact process \cite{s01}, the Ising and $\varphi^4$ models \cite{s07,s15}.

The lace expansion yields a formal recursion equation for the two-point 
function $G_p(x)$, which is similar to the recursion equation for the 
random-walk Green function $S_p(x)$.  For (strictly) SAW, $G_p(x)$ is defined as
\begin{align}\lbeq{SAW-2pt}
G_p(x)=\sum_{\omega:o\to x}p^{|\omega|}\prod_{j=1}^{|\omega|}
 D(\omega_j-\omega_{j-1})\prod_{s<t}(1-\delta_{\omega_s,\omega_t}),
\end{align}
where the sum is over the paths $\omega$ from $o$ to $x$.  The contribution 
from the zero-step walk is regarded as $\delta_{o,x}$.  The last product over 
$s,t$ is either 0 or 1 depending on whether or not $\omega$ intersects to 
itself.  

For Bernoulli bond percolation, in which each bond $\{u,v\}$ is occupied with 
probability $pD(v-u)$ independently of the other bonds, the two-point function 
is defined as 
\begin{align}
G_p(x)=\mP_p(o\conn x),
\end{align}
where $\mP_p$ is the induced law from the above bond-occupation probability 
($p(1-D(o))$ is the expected number of occupied bonds per vertex), and 
$\{o\conn x\}$ is the event that either $x=o$ or there is a self-avoiding path 
of occupied bonds from $o$ to $x$.

For the Ising model, see, e.g., \cite[Section~1.2.4]{csV}.

Due to monotonicity in $p$ and subadditivity in self-avoiding paths, the 
critical point $\pc$ is characterized by the divergence of the susceptibility 
$\chi_p$ for all models, as follows:
\begin{align}
\chi_p=\sum_xG_p(x),&&
\pc=\sup\{p\ge0:\chi_p<\infty\}.
\end{align}

The proof of Theorem~\ref{thm:main} consists of the following two steps:
\begin{description}
\item[Step~1:]
Prove that $G_p(x)$ is bounded by 
$2\lambda\veee{x}_L^{\alpha\wedge2-d}$ if $\alpha\ne2$ and by 
$2\lambda\veee{x}_L^{2-d}/\log\veee{\frac{x}L}_1$ if $\alpha=2$, uniformly in 
$x\in\Zd$ and $p<\pc$, where
\begin{align}\lbeq{lambda}
\lambda=
 \begin{cases}
 \dpst\sup_{x\ne o}S_1(x)\veee{x}_L^{d-\alpha\wedge2}&[\alpha\ne2],\\
 \dpst\sup_{x\ne o}S_1(x)\veee{x}_L^{d-2}\log\veee{\tfrac{x}L}_1\quad
  &[\alpha=2],
 \end{cases}
\end{align}
which is of order $L^{-\alpha\wedge2}$, by Theorem~\ref{theorem:S}.  
\item[Step~2:]
Use the lace expansion as a recursion equation for $G_{\pc}(x)$ to derive 
its asymptotic expression.
\end{description}

To complete \textbf{Step~2} is rather straightforward as soon as 
\textbf{Step~1} is completed; see \cite[Section~3.3]{csIV} for $\alpha\ne2$ 
and \cite[Section~3.5]{csV} for $\alpha=2$.  To complete \textbf{Step~1}, 
it suffices to show that $g_p$, defined as
\begin{align}\lbeq{g-def}
g_p=
 \begin{cases}
 \dpst p\vee\sup_{x\ne o}\frac{G_p(x)}{\lambda\veee{x}_L^{\alpha\wedge2-d}}
  &[\alpha\ne2],\\
 \dpst p\vee\sup_{x\ne o}\frac{G_p(x)}{\lambda\veee{x}_L^{2-d}/\log
  \veee{\frac{x}L}_1}&[\alpha=2],
 \end{cases}
\end{align}
satisfies the following three properties:
\begin{enumerate}[(S1.1)]
\item
$g_1\le1$.
\item[(S1.2)]
$g_p$ is continuous (and nondecreasing) in $p\in[1,\pc)$.
\item[(S1.3)]
$g_p\le3$ implies $g_p\le2$ for every $p\in(1,\pc)$, if $\lambda\ll1$.
\end{enumerate}
The third property implies that there is a prohibited region in the 
$p$--$g_p$ plane.  Therefore, $g_p$ is either $\le2$ or $>3$, as long as 
$p\in(1,\pc)$.  However, due to the continuity (S1.2) with the initial 
condition (S1.1), the possibility of $g_p>3$ is eliminated.  This completes 
\textbf{Step~1}.

(S1.1)--(S1.2) are not so difficult, due to 
\cite[Propositions~3.1--3.3]{csV}.  To show (S1.3), 
we use the lace expansion, which is formally written as
\begin{align}\lbeq{lace-intro}
G_p(x)=\varPi_p(x)+(\varPi_p*pD*G_p)(x),
\end{align}
where (cf., \cite[Section~3.1]{csV})
\begin{align}\lbeq{vPi-def}
\varPi_p(x)=
 \begin{cases}
 \dpst\delta_{o,x}+\sum_{n=1}^\infty\big(-pD(o)\delta+\pi_p\big)^{*n}(x)
  &[\text{SAW}],\\
 \dpst\pi_p(x)+\sum_{n=1}^\infty\big(-pD(o)\big)^n\pi_p^{*(n+1)}(x)\quad
  &[\text{Ising \& percolation}].
 \end{cases}
\end{align}
Here, $\pi_p$ is the alternating series of the nonnegative lace-expansion 
coefficients $\{\pi_p^{\sss(n)}\}_{n=0}^\infty$ ($\pi_p^{\sss(0)}\equiv0$ for 
SAW):
\begin{align}\lbeq{exp-coeff}
\pi_p(x)=\sum_{n=0}^n(-1)^n\pi_p^{\sss(n)}(x).
\end{align}

The proof of Item~(S1.3) goes as follows.  
\begin{enumerate}[(i)]
\item
Bound $\pi_p^{\sss(n)}$ in terms of $G_p$ by using correlation inequalities, 
such as the BK inequality for percolation \cite{bk85}.
\item
Derive an optimal $x$-space bound on $\varPi_p$ in \refeq{vPi-def} by applying 
the hypothesis $g_p\le3$ to the bounds on $\pi_p^{\sss(n)}$ obtained in (i) and 
using convolution bounds (see below) on power functions, with log corrections 
for $\alpha=2$.
\item
Prove the improved bound $g_p\le2$ by applying the bound on $\varPi_p$ 
obtained in (ii) to \refeq{lace-intro}.
\end{enumerate}

From now on, we restrict our attention to percolation.  By the BK inequality, 
the first few terms are bounded as 
\begin{align}
\pi_p^{\sss(0)}(x)\le G_p(x)^2,&&&&
\pi_p^{\sss(1)}(x)\le o\:\raisebox{-7pt}{\includegraphics[scale=0.15]
 {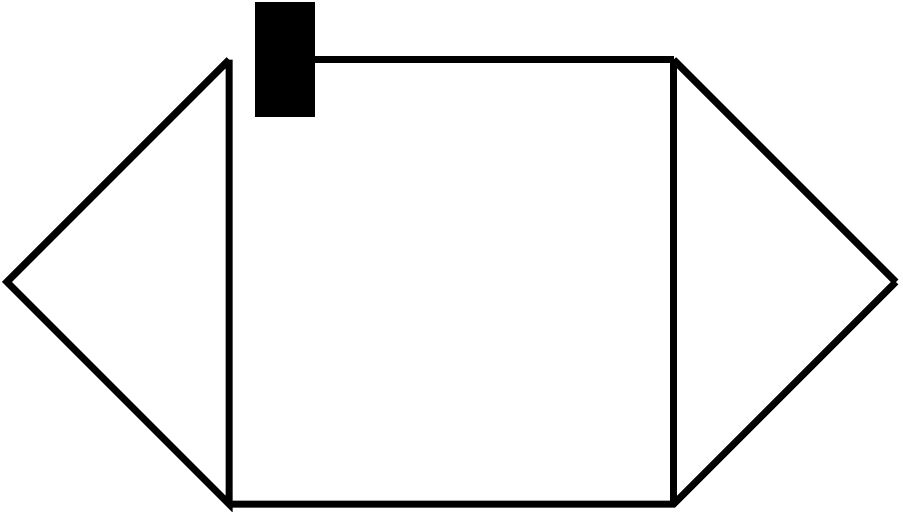}}\,x,&&&&
\pi_p^{\sss(2)}(x)\le o\:\raisebox{-9pt}{\includegraphics[scale=0.15]{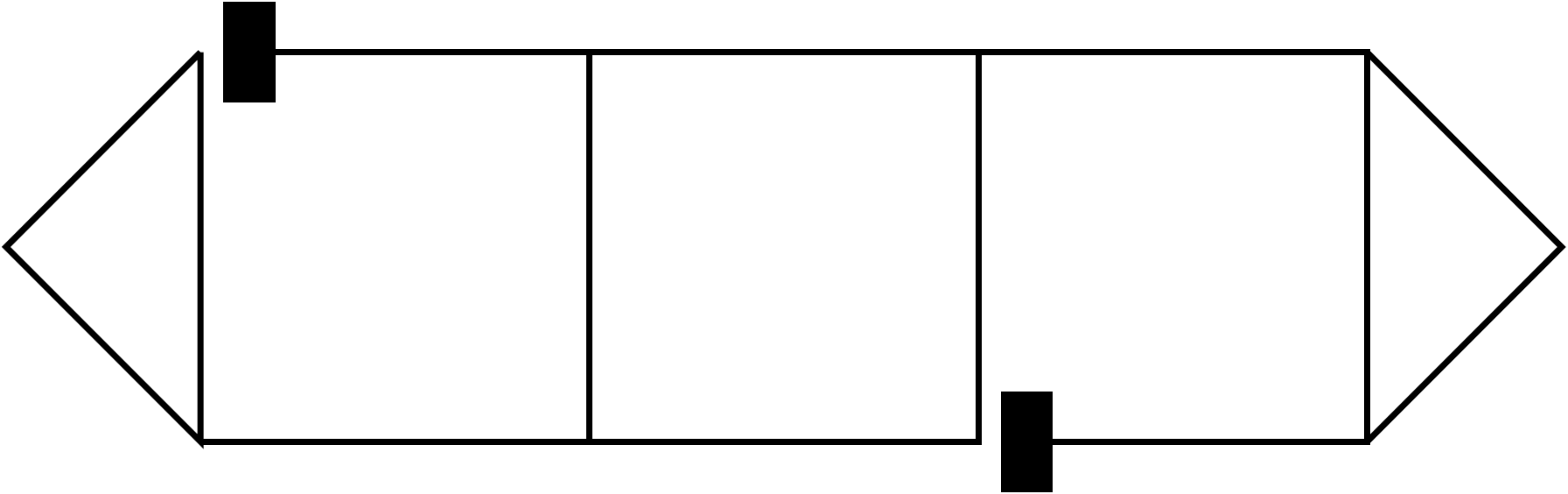}}\,x~
 +\cdots,
\end{align}
where each line segment represents $G_p$, small filled rectangles are $pD$ and 
unlabeled vertices are summed over $\Zd$.  For more explanation on those 
diagrammatic expressions, we refer to the original paper \cite{hs90p}.  Then, 
we use $g_p\le3$ and the following convolution bounds:

\begin{lmm}[Lemma~3.5 of \cite{csV}]\label{lemma:conv-bds}
For $a_1\ge b_1>0$ with $a_1+b_1\ge d$, and for $a_2,b_2\ge0$ with
$a_2\ge b_2$ when $a_1=b_1$, there is an $L$-independent constant
$C=C(d,a_1,a_2,b_1,b_2)<\infty$ such that
\begin{align}\lbeq{conv1}
&\sum_{y\in\Zd}\frac{\veee{x-y}_L^{-a_1}}{(\log\veee{\frac{x-y}L}_1)^{a_2}}
 \frac{\veee{y}_L^{-b_1}}{(\log\veee{\frac{y}L}_1)^{b_2}}\\
&\le\frac{C\,\veee{x}_L^{-b_1}}{(\log\veee{\frac{x}L}_1)^{b_2}}\times
 \begin{cases}
 L^{d-a_1}&[a_1>d],\\
 \log\log\veee{\frac{x}L}_1&[a_1=d,~a_2=1],\\
 (\log\veee{\frac{x}L}_1)^{0\vee(1-a_2)}&[a_1=d,~a_2\ne1],\\
 \veee{x}_L^{d-a_1}&[a_1<d,~a_1+b_1>d],\\
 \veee{x}_L^{b_1}(\log\veee{\frac{x}L}_1)^{0\vee(1-a_2)}&[a_1<d,~a_1+b_1=d,
  ~a_2+b_2>1].
 \end{cases}\nn
\end{align}
\end{lmm}

Take $\pi_p^{\sss(1)}(x)$ for $\alpha=2$, for example.   By repeated 
applications of the above convolution bounds, we can reduce the number of 
vertices (and line segments) one by one, as depicted as follows:
\begin{align}\lbeq{conv-appl}
o\:\raisebox{
 -7pt}{\includegraphics[scale=0.15]{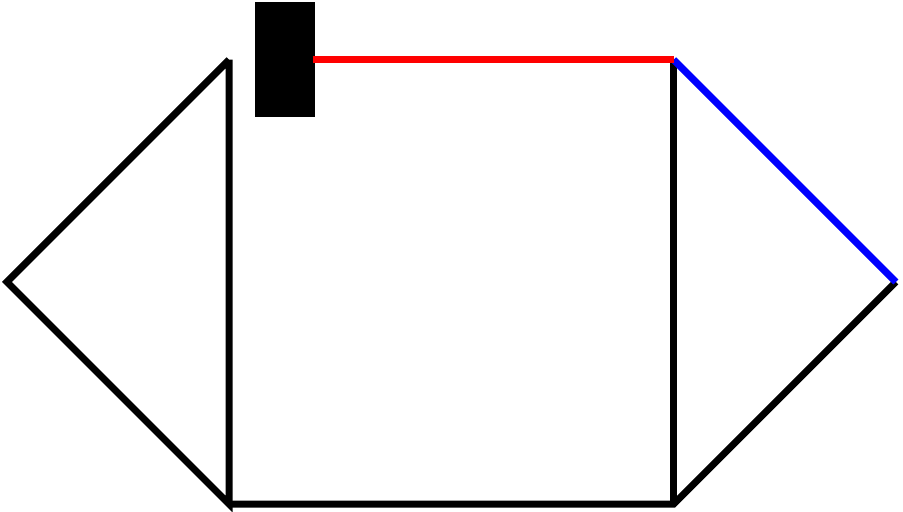}}\,x~\stackrel{\substack{g_p\le3\\
 d\ge4}}\lesssim~o\:\raisebox{-7pt}{\includegraphics[scale=0.15]{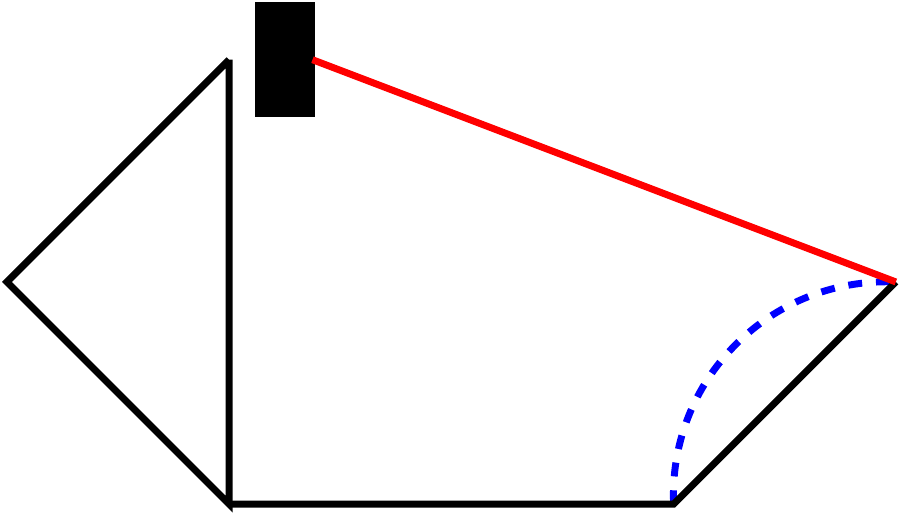}}\,x~+
 ~o\:\raisebox{-7pt}{\includegraphics[scale=0.15]{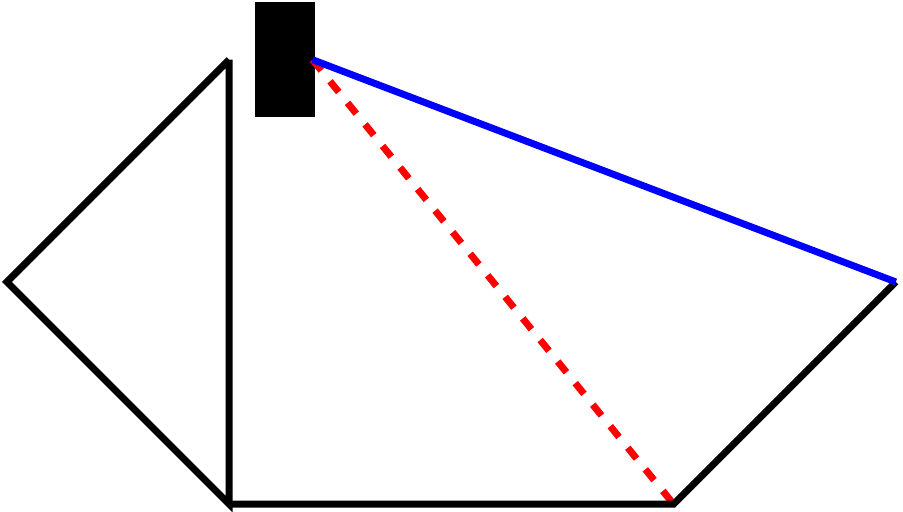}}\,x~\stackrel{
 \substack{g_p\le3\\ d\ge6}}\lesssim~o\:\raisebox{-7pt}{\includegraphics
 [scale=0.15]{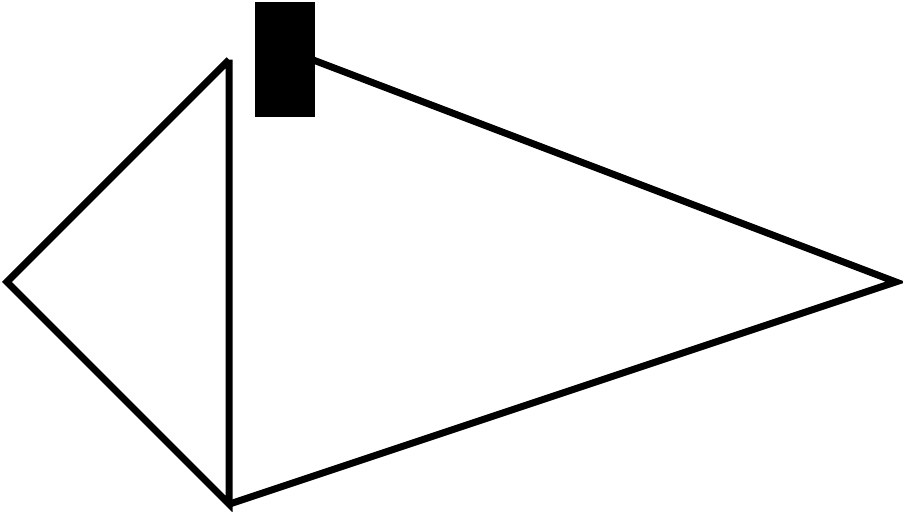}}\,x.
\end{align}

\begin{proof}[Explanation of the above inequality.]
Let $v$ be the unlabeled top-right vertex in the leftmost figure at which three 
line segments (each in red, blue and black) meet, and let $y,z$ be the other 
end vertices of the horizontal (in red) and vertical (in black) line segments, 
respectively.  In the first inequality, we use \refeq{conv1} between the 
vertical line segment and one of the other two line segments, depending on 
whether $|x-v|\ge|y-v|$ or $|x-v|\le|y-v|$.  If $|x-v|\le|y-v|$, then 
$|x-y|\le|x-v|+|y-v|\le2|y-v|$ and therefore
\begin{align}
\sum_{v:|x-v|\le|y-v|}\frac{\veee{x-v}_L^{2-d}}{\log\veee{\frac{x-v}L}_1}
 \frac{\veee{y-v}_L^{2-d}}{\log\veee{\frac{y-v}L}_1}
 \frac{\veee{z-v}_L^{2-d}}{\log\veee{\frac{z-v}L}_1}\nn\\
\le\frac{\veee{\frac{x-y}2}_L^{2-d}}{\log\veee{\frac{x-y}{2L}}_1}
 \sum_v\frac{\veee{x-v}_L^{2-d}}{\log\veee{\frac{x-v}L}_1}
 \frac{\veee{z-v}_L^{2-d}}{\log\veee{\frac{z-v}L}_1}
&\stackrel{d\ge4}\le{}^\exists C'\frac{\veee{x-y}_L^{2-d}}{\log\veee{\frac{x-y}
 L}_1}\underbrace{\frac{\veee{x-z}_L^{4-d}}{\log\veee{\frac{x-z}
 L}_1}}_\text{blue-dotted},
\end{align}
which is depicted as the left figure in the middle expression in 
\refeq{conv-appl}.  Then, by gathering all line segments meeting at $z$ 
(denote the other end vertex of the horizontal line segment by $u$) 
and using \refeq{conv1} again, we obtain
\begin{align}
\sum_z\frac{\veee{x-z}_L^{(4-d)+(2-d)}}{(\log\veee{\frac{x-z}L}_1)^2}
 \frac{\veee{u-z}_L^{2-d}}{\log\veee{\frac{u-z}L}_1}\stackrel{d\ge6}\le
 C\frac{\veee{x-u}_L^{2-d}}{\log\veee{\frac{x-u}L}_1},
\end{align}
which yields the rightmost figure of \refeq{conv-appl}.  We should emphasize 
that the above bound holds even at $\dc=6$, because of the log-squared 
term in the denominator.  This is one of the reasons why the mean-field 
results\footnote{The bubble condition $G_{\pc}^{*2}(o)<\infty$ for SAW/the 
Ising model and the triangle condition $G_{\pc}^{*3}(o)<\infty$ for percolation 
are sufficient conditions for the susceptibility $\chi_p$ and other observables 
to exhibit their mean-field behavior.  The log correction for $\alpha=2$ 
is the key to extend the mean-field results down to $d=\dc$ since, for 
example, the tail of the sum in the triangle condition can be estimated, 
for any $R>1$, as 
\begin{align}
\sum_{x:|x|>R}G_{\pc}(x)\,G_{\pc}^{*2}(x)\stackrel{d\ge4}
 \lesssim\int_R^\infty\frac{\mathrm{d}r}r~\frac{r^{6-d}}{(\log r)^2}\stackrel{d
 \ge6}<\infty.
\end{align}
} 
hold for $d\ge\dc$ (including equality) when $\alpha=2$.

The other case $|x-v|\ge|y-v|$ can be evaluated similarly, and we refrain 
from showing it here.
\end{proof}

Applying the same analysis to the other $\pi_p^{\sss(n)}$ and using 
\refeq{vPi-def}--\refeq{exp-coeff}, we can get (cf., \cite[(3.4)]{csIV} and \cite[(3.29)]{csV})
\begin{align}\lbeq{varPi-bd}
|\varPi_p(x)-\delta_{o,x}|&\le O(L^{-d})\delta_{o,x}+O(\lambda^2)\times
 \begin{cases}
 \veee{x}_L^{(\alpha\wedge2-d)\ell}&[\alpha\ne2],\\
 (\veee{x}_L^{2-d}/\log\veee{\frac{x}L}_1)^\ell\quad&[\alpha=2],
 \end{cases}
\end{align}
where 
\begin{align}\lbeq{elldef}
\ell=
 \begin{cases}
 2&[\text{percolation}],\\
 3\quad&[\text{SAW \& Ising}].
 \end{cases}
\end{align}
Notice from \refeq{varPi-bd} that, if $\alpha<2$ and $d>\dc$ or if $\alpha=2$ 
and $d\ge\dc$, then $\varPi_p*D$ in \refeq{lace-intro} can be treated, after 
normalization, as a probability distribution.  For $\alpha=2$, for example, 
there are finite constants $c,c',c''$ such that
\begin{eqnarray}\lbeq{+ity}
(\varPi_p*D)(x)&\stackrel{\refeq{varPi-bd}}\ge&(1-cL^{-d})D(x)-c'\lambda^2
 \sum_y\frac{\veee{y}_L^{\ell(2-d)}}{(\log\veee{\frac{y}L}_1)^\ell}D(x-y)\nn\\
&\stackrel{\text{Lemma~\ref{lemma:conv-bds}}}\ge&(1-cL^{-d}-c''\lambda^3)D(x),
\end{eqnarray}
which is positive for all $x$, if $\lambda\ll1$.  Therefore, 
\begin{align}\lbeq{new1step}
\cD(x)=\frac{(\varPi_p*D)(x)}{\hat\varPi_p(0)}
\end{align}
is a probability distribution that satisfies all the properties in (D1)--(D3), 
and its Green function $\sum_{n=0}^\infty\cD^{*n}(x)$ is bounded by 
$(1+O(\lambda^3))S_1(x)$ for every $x$ (see \cite[Section~3.2]{csV} for more 
details).  By \refeq{varPi-bd} and Lemma~\ref{lemma:conv-bds}, we obtain 
that, for $x\ne o$,
\begin{align}
G_p(x)\le\big(1+O(\lambda^3)\big)(\varPi_p*S_1)(x)
&\le\big(1+O(\lambda^3)\big)S_1(x)+O(\lambda^4)\frac{\veee{x}_L^{2-d}}
 {\log\veee{\frac{x}L}_1}\nn\\
&\stackrel{\lambda\ll1}\le2\lambda\frac{\veee{x}_L^{2-d}}
 {\log\veee{\frac{x}L}_1},
\end{align}
as required.  This completes all the steps (i)--(iii) for $\alpha\le2$.

If $\alpha>2$, then we can no longer interpret $\varPi_p*D$ as a probability 
distribution, because the second term in \refeq{+ity} decays slower than $D$; 
this is why the model-dependent multiplicative constant $A$ in \refeq{mainasy} 
is reduced to 1 only when $\alpha\le2$.  To overcome this difficulty for 
$\alpha>2$, we assume that the ``derivative" of the $n$-step distribution 
$D^{*n}$ obeys the following bound: for $|y|\le\frac13|x|$,
\begin{align}\lbeq{Dnxdiffbd}
\bigg|D^{*n}(x)-\frac{D^{*n}(x+y)+D^{*n}(x-y)}2\bigg|\le n\,\frac{O(L^{\alpha
 \wedge2})\,\veee{y}_L^2}{\veee{x}_L^{d+\alpha\wedge2+2}}.
\end{align}
We have shown in \cite{csIV} that the compound zeta distribution 
\refeq{comp-zeta} for $\alpha\ne2$ satisfies the above assumption.  
See \cite[Appendix]{csIV} for more details.

\section*{Acknowledgements}
This work was supported in part by JSPS KAKENHI Grant Number 18K03406, 
and in part by JST CREST Grant Number JP22180021.  I would like to thank 
Lung-Chi Chen for the long-time collaboration \cite{csI,csII,csIII,csIV,csV} on 
the long-range models defined by the power-law decaying 
step-distribution/coupling.  I would also like to thank the organizers of 
$17^\text{th}$ International Symposium ``Stochastic Analysis on Large-scale 
Interacting Systems" at RIMS, Kyoto University, during November 5--8, 2018, 
for the opportunity to speak on the topic of this article.  Finally, I would 
like to express my gratitude to the referee for positive feedback and valuable 
suggestions to improve presentation of this paper.

\end{document}